\documentstyle[preprint,aps,floats]{revtex}

\newcommand{\ee}[2]{$#1.\times 10^{-#2}$}
\newcommand{\nn}{\nonumber}

\begin{document}
\draft

\title{Energy and angular momentum flow into a black hole in a binary}
\author{Kashif Alvi}
\address{Physics Division 130-33, California Institute of Technology,
Pasadena, California 91125}
\maketitle
\begin{abstract}
As a black hole in a binary spirals in gradually from large separation,
energy and angular
momentum flow not only to infinity but also into or out of the hole.  In
addition, the hole's horizon area increases slowly during this process.
In this paper, the changes in the black hole's mass, spin, and horizon area
during inspiral are calculated for a hole in a circular binary with a companion
body of possibly comparable mass.  When the binary is composed of equal-mass black holes
that have spins aligned with the orbital angular momentum and are
rapidly rotating (with spins 99.8 percent of their maximal values),
it is found that the fractional increase in the surface area of each
hole's horizon is one percent by the time the binary spirals down to a separation
$b$ of $6M$ (where $M$ is the binary's total mass), and
seven percent down to $b=2M$.  The flow of energy and
angular momentum into the black holes' horizons changes the number of
gravitational-wave cycles in the LIGO band by no more than a tenth of a cycle
by the time the binary reaches $b=2M$.
The results obtained in this paper are relevant for the detection and analysis
of gravitational waves from binary systems containing a black hole.
\end{abstract}

\section{Introduction}

Binary black holes are expected to be among the
primary sources of gravitational waves for interferometric detectors \cite{z&m}.
Since we do not have
exact solutions of Einstein's equations that represent binary black
holes in sufficient generality, we must study these systems
perturbatively and/or numerically.  One regime in which the evolution of
binary black holes is well understood is the early inspiral phase.  In this
phase, the holes' separation is still much larger than the
binary's total mass, and post-Newtonian expansions
can be used to analyze the system.
Eventually radiation reaction drives the holes together
and the post-Newtonian approximation fails.  The binary's subsequent evolution
must be studied numerically.

While the flow of energy and angular momentum to infinity during inspiral has been
calculated to high post-Newtonian order, to date the flow into or out of the
black holes' horizons has not been computed except in the
extreme-mass-ratio limit; and in that limit, it has been done
to very high post-Newtonian order \cite{tmt}
(for numerical work, see e.g.\ \cite{hughes}).
Absorption (or emission) of energy and angular momentum by the holes' horizons,
while much smaller than emission to infinity, might still be important
because extraction of weak gravitational signals from noisy detector output using
matched filtering requires
knowledge of the orbital evolution to very high accuracy, and black hole
absorption/emission might affect the evolution at that level.
Two purposes of this paper are to
calculate black hole absorption/emission of energy and angular momentum to
leading order in a circular binary with holes of possibly comparable mass, and to
investigate whether it is relevant for detection and analysis of
gravitational waves.

A third purpose of this paper is to provide some information on the
interface between the inspiral and merger phases of binary evolution.
Numerical simulations of binary black holes typically begin computing at this
interface and need initial data representing holes that have
spiraled in from infinity, i.e.\ initial data tied to the inspiral phase and to the
post-Newtonian expansions used to describe it.  One approach to obtaining such initial
data is given in \cite{alvi}.  Since initial data of this sort are not yet being used,
one needs to relate the masses, spins, and horizon
areas of the black holes present in currently used initial data to the corresponding
quantities when the holes were infinitely separated.  For this purpose, it
is necessary to know how these quantities change during inspiral.  In
this paper, I calculate the leading-order changes in the holes' masses,
spins, and horizon areas during inspiral for a circular binary.

Recently, Price and Whelan \cite{p&w} have emphasized the role of angular momentum
absorption/emission by rapidly rotating black holes at the end of inspiral, when
the holes are beginning to merge.  Here I focus on the earlier stages of inspiral, when
the holes are widely (or moderately) separated and their
gravitational effects on each other can be
described using black hole perturbation theory.

The results obtained in this paper are actually valid
for a black hole in a binary with any companion
body (e.g.\ a neutron star) that is well separated from the hole.  The formulas
for the changes in black hole quantities presented here depend only on the companion
body's mass and not on its internal structure.  These formulas therefore remain
valid when the companion is not a black hole.

\section{Framework}

I follow the field-theory-in-flat-spacetime notation used in the
literature on post-Newtonian
expansions (e.g.\ \cite{bdiww,kidder}) and denote 3-vectors by bold-face letters.
A centered dot between 3-vectors denotes the usual inner product
in flat 3-space; a hatted 3-vector represents the unit vector in that direction.

Consider a black hole binary undergoing circular motion with the separation
$b$ between the holes much larger than their total mass $M=M_1+M_2$, where
$M_B$ denotes the mass of the $B$th hole.  Define $\mu=M_1 M_2/M$ and
$\eta=\mu/M$.  Label the holes BH1 and BH2, and
denote their spins by ${\bf S}_B$ and horizon areas by $A_B$ for $B=1,2$.
Let $S_B=({\bf S}_B\cdot{\bf S}_B)^{1/2}$ be the spin magnitudes,
and define the parameters $\chi_B$ by $S_B=\chi_B M_B^2$ ($B=1,2$).
Throughout this paper I assume $\chi_B\leq 0.998$ ($B=1,2$); this restriction
is based on the analysis in \cite{kip}.

Define
each black hole's horizon radius $r_{HB}=M_B[1+(1-\chi_B^2)^{1/2}]$, angular velocity
$\Omega_{HB}=\chi_B(2r_{HB})^{-1}$, and surface gravity
$\kappa_B=(1-\chi_B^2)^{1/2}(2r_{HB})^{-1}$ ($B=1,2$).
Introduce the following Newtonian quantities for the binary: the orbital
angular momentum ${\bf L}_N$, the orbital angular velocity $\Omega_N
=(M/b^3)^{1/2}$, and the relative velocity $v=(M/b)^{1/2}$.  By assumption,
$v\ll 1$.

Since the black holes are widely separated, each hole has a surrounding region
that satisfies the following properties: (i) it is far enough from the hole
that gravity is weak there; (ii) it does not extend so far that the
companion hole's tidal field varies appreciably in
the region \cite{t&h}.  We can place in this region an inertial coordinate system in
which the hole is (momentarily) at rest.  This region and its local
coordinates are referred to as the black hole's local asymptotic rest frame
(LARF) \cite{t&h}.  Label the two regions around the holes LARF1 and LARF2.

Usually mass and angular momentum are only defined globally in general
relativity, using fields at infinity, since precise local definitions are
not available.  However, for a black hole well separated from its companion,
one can define the
hole's mass and angular momentum using fields measured in the hole's
LARF; these definitions are inherently ambiguous \cite{t&h,purdue,b&c,favata}.
(For further discussion of the ambiguities, see Sec.~\ref{sec:discussion}.)
I refer to these definitions when discussing
a black hole's mass and angular momentum in this paper.
I calculate the rates of change of these quantities
as measured in the LARF---that is, with respect to time $t$
measured by an inertial observer in the LARF.  When integrated over the duration
of inspiral, these rates of change should give results exceeding
the ambiguities in the definitions of mass and angular momentum, in order to be
relevant to the analysis of initial data at the interface between inspiral and merger.
This issue will be discussed further in Sec.~\ref{sec:discussion}.

I also consider slices of constant time $t$ that begin in the LARF and
extend into the black hole, intersecting the horizon in 2-surfaces that
correspond to constant ingoing-time slices of a Kerr black hole's horizon.
(Alternatively, one can consider slices that intersect a ``stretched horizon''
as discussed in \cite{membrane} and references therein.)
The rate of area increase of these 2-surfaces can be calculated using the
results of Hawking and Hartle \cite{h&h} combined with black hole perturbation
theory \cite{teuk,t&p}.
The quantities $dM_B/dt$ and $dS_B/dt$ can then be obtained from $dA_B/dt$
using the first law of black hole mechanics $dM=(\kappa/8\pi)dA
+\Omega_H dJ$ and the relation $\omega dJ=mdM$ for black hole perturbation modes of
angular frequency $\omega$ and azimuthal angular number $m$ \cite{t&p,carter,membrane}.
(Here $J$ refers to the black hole's angular momentum.)

Throughout this paper, I focus on BH1 and the changes in its parameters.
The corresponding formulas for BH2 are simply obtained by exchanging the subscripts
$1\leftrightarrow 2$ in the final results [e.g.\ Eqs.~(\ref{eq:alignedspins})].
In Sec.~\ref{sec:stationary}, I consider the special situation in which BH2
is held stationary with respect to BH1.  The results from this artificial
scenario are used in Sec.~\ref{sec:equatorial} to analyze a circular binary
with black hole spins aligned
or anti-aligned with ${\bf L}_N$.  The more general case of spins not fully aligned
or anti-aligned with ${\bf L}_N$ is treated in Sec.~\ref{sec:nonequatorial}.

\section{Stationary companion}
\label{sec:stationary}

In this section, I calculate the tidal distortion BH1 suffers when
BH2 is held stationary.  This involves solving for the Weyl tensor component
$\psi_0$, which contains complete information about the gravitational perturbation
on BH1, using the Teukolsky formalism \cite{teuk}.  With $\psi_0$ in hand, the
rates of change of BH1 parameters can be calculated using the results of Hawking
and Hartle \cite{h&h} and Teukolsky and Press \cite{t&p}.

The first step in this process is to calculate BH2's tidal field as seen in
LARF1.  I will consider only the lowest-order Newtonian tidal field, which is
approximately constant throughout LARF1.  To calculate this field
and its effect on BH1, consider
first a fictitious Euclidean 3-space containing a single stationary body of mass
$M_2$ at coordinate location $(b,\theta_0,\phi_0)$ in a spherical polar coordinate
system.  The Newtonian gravitational potential at
the field point $(r,\theta,\phi)$ is given in these coordinates by
\begin{equation}
\Phi(r,\theta,\phi) = -4\pi {M_2\over b}\sum_{l=0}^{\infty}\sum_{m=-l}^{l}
	(2l+1)^{-1}\left({r\over b}\right)^l Y_{lm}^{*}(\theta_0,\phi_0)
	Y_{lm}(\theta,\phi)
\end{equation}
for $r<b$.

We are interested in the gravitational field only in a small neighborhood of
the origin satisfying $r\ll b$.  In particular, we would like to evaluate the
body's tidal field at the origin, so only the $l=2$ part $\Phi^{(2)}$ of
$\Phi$ is relevant.  The (electric-type) tidal field is given by ${\cal E}_{ij}=
\Phi_{,ij}^{(2)}$ in Cartesian coordinates.  Taking these derivatives in
spherical coordinates and evaluating in the usual spherical orthonormal basis
yields the tidal field components ${\cal E}_{\hat{\theta}\hat{\theta}}$,
${\cal E}_{\hat{\theta}\hat{\phi}}$, ${\cal E}_{\hat{\phi}\hat{\phi}}$
near the origin $r=0$.  The particular combination of relevance to us
(see below) is in this way determined to be
\begin{equation}
{\cal E}_{\hat{\phi}\hat{\phi}}-{\cal E}_{\hat{\theta}\hat{\theta}}
	-2i{\cal E}_{\hat{\theta}\hat{\phi}} = {8\pi\sqrt{6}M_2\over 5b^3}
	\sum_{m=-2}^{2}{}_2 Y_{2m}(\theta,\phi) Y_{2m}^{*}(\theta_0,\phi_0).
\label{eq:psi0form}
\end{equation}
Here the functions $_{2}Y_{2m}(\theta,\phi)$
are spin-weighted spherical harmonics \cite{goldberg}.

Return now to the black hole binary.  The region near BH1, including LARF1, can
be described as a perturbed Kerr black hole, and so can be covered by
Boyer-Lindquist coordinates $(t,r,\theta,\phi)$.
We would like to solve the Teukolsky equation \cite{teuk} in this region for
the Weyl tensor component $\psi_0(r,\theta,\phi)$.
If we were considering a single perturbed Kerr black hole as the entire spacetime,
the asymptotic form
of $\psi_0$ as $r/M_1\rightarrow \infty$ would be the combination
${\cal E}_{\hat{\phi}\hat{\phi}}-{\cal E}_{\hat{\theta}\hat{\theta}}
-2i{\cal E}_{\hat{\theta}\hat{\phi}}$ of the external tidal field \cite{membrane},
since $\psi_0$ vanishes for an unperturbed black hole.
In our binary system, $\psi_0$ acquires this asymptotic form for
$M_1 \ll r \ll b$, i.e.\ in LARF1, with the tidal field ${\cal E}_{ij}$ being
that of BH2.  To lowest order, this tidal field is exactly the Newtonian
field of a body of mass $M_2$ at separation $b$, which was calculated above;
in particular, the angular dependence of $\psi_0$ in LARF1 is given by
Eq.~(\ref{eq:psi0form}), but with $\theta$ and $\phi$ now representing
Boyer-Lindquist coordinates, and $\theta_0$ and $\phi_0$ now representing BH2's
angular coordinates as seen in LARF1.  Therefore, to solve for the perturbation
$\psi_0$ on BH1, we impose the LARF1 boundary condition
\begin{equation}
\psi_0 \rightarrow {8\pi\sqrt{6}M_2\over 5b^3}\sum_{m=-2}^{2}
	{}_2 Y_{2m}(\theta,\phi) Y_{2m}^{*}(\theta_0,\phi_0)
\label{eq:larfbc}
\end{equation}
for $M_1 \ll r \ll b$.

It now remains to solve the
Teukolsky equation for $\psi_0$ with the boundary
condition~(\ref{eq:larfbc}) and
an appropriate no-outgoing-wave boundary condition at the black hole
horizon \cite{teuk}.  We express $\psi_0$ as a sum of modes
\begin{equation}
\psi_0 = \sum_{m=-2}^2 {}_2 Y_{2m}(\theta,\phi) R_m(r)
\label{eq:modesum}
\end{equation}
and solve the radial
Teukolsky equation for $R_m(r)$ subject to the no-outgoing-wave
boundary condition at the horizon.
This yields the radial functions (Eq.~(5.7) in Ch.~VI of \cite{thesis})
\begin{equation}
R_m(r) = C_m x^{\gamma_m-2}(1+x)^{-\gamma_m-2}F(-4,1,-1+2\gamma_m,-x)
\end{equation}
for $m\neq 0$.  Here
\begin{equation}
\gamma_m = \frac{im\chi_1}{2(1-\chi_1^2)^{1/2}}, \qquad x = \frac{r-r_{H1}}
	{2M_1(1-\chi_1^2)^{1/2}},
\end{equation}
and $F$ is a hypergeometric function.  The $m=0$ mode can be
treated separately; since a full treatment reveals that this mode does not
contribute to the rates
of change of black hole parameters, I ignore it here.  The constants $C_m$ are
determined by imposing the LARF1
boundary condition~(\ref{eq:larfbc}); we obtain
\begin{equation}
C_m = \frac{8\pi M_2}{5b^3\sqrt{6}}\gamma_m(\gamma_m+1)(4\gamma^2_m-1)
	Y_{2m}^{*}(\theta_0,\phi_0).
\end{equation}

The leading-order tidal distortion of BH1 due to the presence
of a stationary companion of mass $M_2$ has now been determined.
This information allows us to calculate the rates of change of BH1
quantities using the results of Hawking and Hartle \cite{h&h}.  In fact, given the
modal decomposition~(\ref{eq:modesum}), we can easily obtain the relevant
rates using explicit formulas provided by Teukolsky
and Press \cite{t&p}.  The results are $dM_1/dt = 0$ and
\begin{eqnarray}
{dA_1\over dt} &=& \frac{64\pi M_1^5 M_2^2\chi_1^2\sin^2\theta_0}
	{5b^6(1-\chi_1^2)^{1/2}}\left(1-{3\over 4}\chi_1^2+{15\over 4}
	\chi_1^2\sin^2\theta_0\right),\nn\\
{dS_1\over dt} &=& -\frac{(1-\chi_1^2)^{1/2}}{8\pi\chi_1}{dA_1\over dt}
	= -\frac{8 M_1^5 M_2^2}{5b^6}\chi_1\sin^2\theta_0
	\left(1-{3\over 4}\chi_1^2+{15\over 4}
	\chi_1^2\sin^2\theta_0\right).\nn\\
\label{eq:stationary}
\end{eqnarray}
Here $\theta_0$ is BH2's $\theta$-coordinate---that is, its polar angle with respect to
${\bf S}_1$---as measured in LARF1.

Since the effects of only the leading-order tidal field were taken into
account above, the expressions~(\ref{eq:stationary}) are actually valid for
any companion body of mass $M_2$, not just a black hole.  The
rates~(\ref{eq:stationary}) of area increase and spin-down have already been
derived by Teukolsky \cite{thesis} in the extreme-mass-ratio limit,
i.e.\ for $M_2\ll M_1$.
The derivation I have presented above establishes the validity of the
expressions~(\ref{eq:stationary}) for comparable-mass black holes
as well.  Hartle and collaborators \cite{h&h,hartle1,hartle2,membrane} have shown
that the spin-down of a black hole by an
external tidal field is analogous to the Newtonian tidal friction process in
a planet-moon system.

The results~(\ref{eq:stationary}) will be used in the next sections to
obtain the corresponding
formulas for a binary undergoing circular motion.

\section{Equatorial orbits}
\label{sec:equatorial}

In this section I study special configurations of the binary in which
the black holes are in a circular orbit and have their spins
aligned or anti-aligned with the orbital angular momentum ${\bf L}_N$.
In these scenarios there is no precession of the
angular momenta: the spins remain aligned or anti-aligned with ${\bf L}_N$.
As a result, the companion to each of the
holes orbits in the hole's equatorial plane; more precisely, the external tidal
field seen by each of the holes rigidly rotates about an axis
parallel or antiparallel to the hole's spin axis.

In Boyer-Lindquist coordinates $(t,r,\theta,\phi)$ centered on BH1,
with ${\bf S}_1$ along $\theta=0$, the $t$- and $\phi$-dependence
of the companion's tidal field enter in the combination
$\phi - \Omega t$.  The rotation rate $\Omega$ of the tidal field
as seen in LARF1 is to leading order
$\Omega=(\hat{\bf L}_N\cdot\hat{\bf S}_1)\Omega_N$, where $\hat{\bf L}_N\cdot
\hat{\bf S}_1=+1$ ($-1$) for a prograde (retrograde) orbit.
The first correction to this expression for $\Omega$ is $O(v^2)$ higher
(see Eq.~(3.12) in \cite{alvi}), and will be ignored in this paper.

\subsection{Instantaneous rates}

In the rigid $\phi$-rotation case, simple formulas given in Eqs.~(7.21) of
\cite{membrane} (and reproduced below) specify the rates of change of
black hole quantities in terms of a horizon integral $I$ that depends on
the particular perturbing gravitational fields present:
\begin{eqnarray}
{dS_1\over dt} &=& (\Omega-\Omega_{H1})I,
	\qquad {dM_1\over dt} = \Omega{dS_1\over dt} =
	\Omega(\Omega-\Omega_{H1})I,\nn\\
\frac{\kappa_1}{8\pi}{dA_1\over dt} &=&
	(\Omega-\Omega_{H1}){dS_1\over dt} =
	(\Omega-\Omega_{H1})^2 I.\nn\\
\label{eq:rigidrot}
\end{eqnarray}
In terms of ingoing Kerr coordinates $(V,r,\theta,\tilde{\phi})$ (see e.g.\ \cite{mtw}
for a definition), 
$I$ is an integral of a function of $\theta$ and
($\tilde{\phi}-\Omega V$) over a constant-$V$ slice of the horizon.
Since $\tilde{\phi}$-rotations are isometries of the horizon
metric, $I$ is independent of $V$.

Consider an expansion of $I$ in powers
of $M_1\Omega$, which is $O(v^3)$ and hence much smaller than 1.
The zeroth-order part $I_0=\left.I\right|_{\Omega=0}$
is independent of $\Omega$ and, in our situation of binary black holes,
can be easily obtained from the results
for a stationary companion.  From Eqs.~(\ref{eq:rigidrot}), we have
$\left.\dot{S}_1\right|_{\Omega=0}=-\Omega_{H1}I_0$, where an overdot
indicates a time derivative.  But $\Omega=0$ corresponds to
a stationary companion, and in this case we have an explicit expression for $\dot{S}_1$
in Eqs.~(\ref{eq:stationary}).  Equating $\dot{S}_1$ in Eqs.~(\ref{eq:stationary})
to $-\Omega_{H1}I_0$ yields
\begin{equation}
I_0(\theta_0) = \frac{16r_{H1}}{5b^6}M_1^5 M_2^2\sin^2\theta_0
	\left(1-{3\over 4}\chi_1^2 + {15\over 4}\chi_1^2
	\sin^2\theta_0\right),
\label{eq:I0}
\end{equation}
where $\theta_0=\pi/2$ for the equatorial orbits considered in
this section.  The general expression~(\ref{eq:I0}) with
a wider range of values for $\theta_0$ will be used for non-equatorial orbits
in the next section.  Since the first correction to $I_0$ in the expansion of
$I$ in powers of $M_1\Omega$ is $O(M_1\Omega)=O(v^3)$, I will approximate $I$ by $I_0$
throughout this paper.

Note that Eqs.~(\ref{eq:rigidrot}) are, strictly speaking, valid only for constant
rotation rates $\Omega$.  In our situation, radiation reaction drives the binary
together and so $\Omega$ changes during inspiral.  However, the timescale for these
changes is the inspiral timescale $\tau_{\rm ins}\sim bv^{-6}$, where ``$\sim$'' means
``is of the order of''; this is to be compared to
the timescale $\kappa_1^{-1}$ on which the divergence and shear of the null
generators of the horizon probe the future \cite{h&h,carter,membrane}.
By assumption, $\chi_1$ is less than or equal to $0.998$; this implies
$\kappa_1^{-1}<34M_1$, so $\kappa_1^{-1}$ is much smaller than $\tau_{\rm ins}$.
Therefore Eqs.~(\ref{eq:rigidrot}) are valid in our binary system to a very good
approximation.  The various timescales of interest to us will be discussed in
more detail below.

Note also that Eqs.~(\ref{eq:rigidrot}) [and, in addition, Eqs.~(\ref{eq:alignedspins}),
(\ref{eq:nonequatorial}), and (\ref{eq:orbitaverages}) below]
are valid only when integrated over time
intervals much longer than $\kappa_1^{-1}$ (see the discussion in Sec.~VI.C.11
of \cite{membrane}).  In this paper, I am interested in integrating
these equations over the entire inspiral---that is, over time intervals of order
$\tau_{\rm ins}$---so this condition is certainly satisfied.

After putting $I_0(\pi/2)$ and $\Omega=(\hat{\bf L}_N\cdot\hat{\bf S}_1)\Omega_N$
into Eqs.~(\ref{eq:rigidrot}), we obtain the following
rates of change of BH1 quantities for a circular
orbit with spins aligned or anti-aligned with ${\bf L}_N$:
\begin{eqnarray}
{dS_1\over dt} &=& (\Omega-\Omega_{H1})I_0(\pi/2)\nn\\
	&=& \left({dJ\over dt}\right)_N {v^5\over 4}
	\left(M_1\over M\right)^3(1+3\chi_1^2)\left\{
	-\chi_1+2(\hat{\bf L}_N\cdot\hat{\bf S}_1)
	\left[1+(1-\chi_1^2)^{1/2}\right]{M_1\over M}v^3\right\},\nn\\
{dM_1\over dt} &=& \Omega(\Omega-\Omega_{H1})I_0(\pi/2)\nn\\
	&=& \left({dE\over dt}\right)_N {v^5\over 4}
	\left(M_1\over M\right)^3(1+3\chi_1^2)\left\{
	-(\hat{\bf L}_N\cdot\hat{\bf S}_1)\chi_1+
	2\left[1+(1-\chi_1^2)^{1/2}\right]{M_1\over M}v^3\right\},\nn\\
{dA_1\over dt} &=& 8\pi \kappa_1^{-1}(\Omega-\Omega_{H1})^2 I_0(\pi/2)\nn\\
	&=& \frac{64\pi M_1^5 M_2^2(1+3\chi_1^2)}{5b^6(1-\chi_1^2)^{1/2}}
	\left\{\chi_1-2(\hat{\bf L}_N\cdot\hat{\bf S}_1)
	\left[1+(1-\chi_1^2)^{1/2}\right]{M_1\over M}v^3\right\}^2.\nn\\
\label{eq:alignedspins}
\end{eqnarray}
In these formulas, the Newtonian quadrupole expressions for energy and
angular momentum flow to infinity are \cite{p&m,peters}
\begin{equation}
\left({dE\over dt}\right)_N = {32\over 5}\eta^2 v^{10}, \qquad
	\left({dJ\over dt}\right)_N = {32\over 5}\eta^2 Mv^7,
\label{eq:quadrupole}
\end{equation}
where $v=(M/b)^{1/2}$ and $\eta=M_1 M_2/M^2$.  Note that energy and angular momentum
absorption/emission
by a rotating (non-rotating) black hole is 2.5 (4) post-Newtonian orders below the quadrupole
emission~(\ref{eq:quadrupole}) to infinity, as first derived in the extreme-mass-ratio
limit by Poisson and Sasaki \cite{p&s} and Tagoshi, Mano, and Takasugi \cite{tmt}.
The rates of change for BH2 are obtained by exchanging the subscripts
$1\leftrightarrow 2$ in the formulas~(\ref{eq:alignedspins}).

The energy absorption/emission rate $\dot{M}_1$ given above agrees in the limit
$M_2/M\rightarrow 0$ with the lowest-order expression obtained by
Tagoshi, Mano, and Takasugi \cite{tmt}.  Those authors have calculated
this rate in the extreme-mass-ratio limit, for a circular equatorial orbit,
to much higher order in $v$
than I have done here.  However, their results are not applicable to
comparable-mass binaries, while the formulas~(\ref{eq:alignedspins}) are.

The expressions~(\ref{eq:alignedspins}) are valid even if BH1's companion
is not a black hole, provided the companion's mass is substituted for $M_2$.

\subsection{Total changes during inspiral}

In this subsection, I integrate Eqs.~(\ref{eq:alignedspins}) to calculate
the total changes in $M_1$, $S_1$, and $A_1$
during inspiral.  I take into account only the leading-order
Newtonian effects of radiation reaction when computing orbital decay; given
this approximation, the orbital separation $b$ evolves as \cite{peters,mtw}
\begin{equation}
b(t) = b_0(1-t/\tau_0)^{1/4},
\label{eq:inspiral}
\end{equation}
where $\tau_0=(5/256)b_0^4(\mu M^2)^{-1}$.  I also
ignore all post-Newtonian corrections to the orbital angular velocity $\Omega_N$.

It is convenient to parametrize the orbit by separation $b$ instead of time $t$.
The total change
in a parameter, say $S_1$, from infinite separation to separation $b$ is denoted
$\Delta S_1(b)$ and is calculated by integrating Eqs.~(\ref{eq:alignedspins}).
As a first approximation, the quantities
$M_B$ and $S_B$ ($B=1,2$) on the right-hand sides of
Eqs.~(\ref{eq:alignedspins}) can be considered constants during
inspiral.  The reason is that the timescales for evolution of $M_B$ and $S_B$
are much longer than the inspiral timescale $\tau_{\rm ins}\sim bv^{-6}$.  Indeed,
the timescale for evolution of the masses is
$\tau_{M}\sim M_B/\dot{M}_B\sim bv^{-13}$, and for the spins is $\tau_{S}\sim S_B/
\dot{S}_B\sim bv^{-10}$.
So $\tau_M\gg\tau_S\gg\tau_{\rm ins}$ and we can safely
treat $M_B$ and $S_B$ ($B=1,2$) as constants on the right-hand sides of
Eqs.~(\ref{eq:alignedspins}) when integrating over inspiral.

With these approximations, the normalized changes in BH1 parameters from
infinite separation to separation $b$ are
\begin{eqnarray}
\frac{\Delta S_1}{M_1^2}(b) &=& {\eta M_1\over 4M}(1+3\chi_1^2)
	\left\{-{\chi_1\over 4}\left({M\over b}\right)^2
	+(\hat{\bf L}_N\cdot\hat{\bf S}_1)\left[1+(1-\chi_1^2)^{1/2}\right]
	{2M_1\over 7M}\left({M\over b}\right)^{7/2}\right\},\nn\\
\frac{\Delta M_1}{M_1}(b) &=& {\eta\over 4}\left({M_1\over M}\right)^2(1+3\chi_1^2)
	\left\{-(\hat{\bf L}_N\cdot\hat{\bf S}_1){\chi_1\over 7}
	\left({M\over b}\right)^{7/2}+\left[1+(1-\chi_1^2)^{1/2}\right]{M_1\over 5M}
	\left({M\over b}\right)^{5}\right\},\nn\\
\frac{\Delta A_1}{A_1}(b) &=& \frac{\eta M_1^2(1+3\chi_1^2)}{8Mr_{H1}(1-\chi_1^2)^{1/2}}
	\Biggl[{\chi_1^2\over 2}\left({M\over b}\right)^2
	-(\hat{\bf L}_N\cdot\hat{\bf S}_1){8\chi_1\over 7}
	{r_{H1}\over M}\left({M\over b}\right)^{7/2}\nn\\
	&&\mbox{} +{4\over 5}
	\left({r_{H1}\over M}\right)^2\left({M\over b}\right)^5\Biggr],\nn\\
\label{eq:changes}
\end{eqnarray}
where $r_{H1}=M_1[1+(1-\chi_1^2)^{1/2}]$.
To evaluate these changes, one can put into the formulas~(\ref{eq:changes})
the values of $M_1$, $S_1$,
and $A_1$ at infinite separation or, for that matter, at any separation much larger
than $M$, because the changes in these quantities during inspiral are small.
Once again, the changes for BH2 are obtained by exchanging the subscripts
$1\leftrightarrow 2$ in the expressions~(\ref{eq:changes}).

The normalized parameter changes~(\ref{eq:changes}),
evaluated at different stages during inspiral,
are displayed in Tables~\ref{table:spin}-\ref{table:area}
for an equal-mass binary ($M_1=M_2$) with
$\hat{\bf L}_N\cdot\hat{\bf S}_1=1$.  Since a binary composed of slowly rotating
black holes is expected to be undergoing a transition from inspiral to merger
by the time it reaches $b=6M$, the endpoint of integration
is chosen to be $b=6M$ when $\chi_1=0$ and $0.5$.
For rapidly rotating holes ($\chi_1=\chi_2=0.998$), the
endpoint is chosen to be $b=2M$.  The assumption $M\ll b$ is not valid at and near these
endpoints.  The results presented here are most accurate in the early stages of
inspiral, when the black holes are widely separated, and are a rough
estimate of the true parameter changes in the late stages of inspiral.

\begin{table}
\begin{center}
\begin{tabular}{ccccc}
$\chi_1$ & $b/M=100$ & $b/M=20$ & $b/M=6$ & $b/M=2$ \\ \hline
0 & \ee{9}{10} & \ee{2}{7} & \ee{2}{5} &\\
0.5 & \ee{-7}{7} & \ee{-2}{5} & \ee{-2}{4} &\\
0.998 & \ee{-3}{6} & \ee{-8}{5} & \ee{-8}{4} & \ee{-6}{3} \\
\end{tabular}
\end{center}
\caption{
Normalized change $\Delta S_1/M_1^2$ in spin evaluated at $b/M$=100, 20, and 6
for an equal-mass binary with $\hat{\bf L}_N\cdot\hat{\bf S}_1=1$.
For rapidly rotating holes ($\chi_1=\chi_2=0.998$), this change is also evaluated
at $b/M=2$.}
\label{table:spin}
\end{table}

\begin{table}
\begin{center}
\begin{tabular}{ccccc}
$\chi_1$ & $b/M=100$ & $b/M=20$ & $b/M=6$ & $b/M=2$ \\ \hline
0 & \ee{3}{13} & \ee{1}{9} & \ee{4}{7} &\\
0.5 & \ee{-2}{10} & \ee{-5}{8} & \ee{-3}{6} &\\
0.998 & \ee{-9}{10} & \ee{-2}{7} & \ee{-2}{5} & \ee{-6}{4} \\
\end{tabular}
\end{center}
\caption{
Normalized change $\Delta M_1/M_1$ in mass evaluated at $b/M$=100, 20, and 6
for an equal-mass binary with $\hat{\bf L}_N\cdot\hat{\bf S}_1=1$.
For rapidly rotating holes ($\chi_1=\chi_2=0.998$), this change is also evaluated
at $b/M=2$.}
\label{table:mass}
\end{table}

\begin{table}
\begin{center}
\begin{tabular}{ccccc}
$\chi_1$ & $b/M=100$ & $b/M=20$ & $b/M=6$ & $b/M=2$ \\ \hline
0 & \ee{6}{13} & \ee{2}{9} & \ee{8}{7} &\\
0.5 & \ee{2}{7} & \ee{5}{6} & \ee{4}{5} &\\
0.998 & \ee{5}{5} & \ee{1}{3} & \ee{1}{2} & \ee{7}{2} \\
\end{tabular}
\end{center}
\caption{
Normalized change $\Delta A_1/A_1$ in horizon area evaluated at $b/M$=100, 20, and 6
for an equal-mass binary with $\hat{\bf L}_N\cdot\hat{\bf S}_1=1$.
For rapidly rotating holes ($\chi_1=\chi_2=0.998$), this change is also evaluated
at $b/M=2$.}
\label{table:area}
\end{table}

\subsection{Effect on orbital evolution}

The orbital evolution of binary black holes is affected by the absorption/emission
of energy and angular momentum by the holes.  In particular, the number of orbits---and
hence the number of gravitational-wave cycles emitted to infinity---changes when
black hole absorption/emission is accounted for.  To estimate this effect,
let us consider a circular, nearly Newtonian binary, with spins aligned or anti-aligned
with ${\bf L}_N$, that is losing orbital energy and angular
momentum to infinity via Newtonian quadrupole radiation~(\ref{eq:quadrupole}), and
to the black holes via tidal interaction as specified by Eqs.~(\ref{eq:alignedspins}).
Since $\dot{M}_B = \Omega\dot{S}_B = (\hat{\bf L}_N\cdot\hat{\bf S}_B)\Omega_N\dot{S}_B$
($B=1,2$), circular, nearly Newtonian orbits remain circular.
Therefore the evolution of the separation $b(t)$ is determined by setting the
rate of change of Newtonian orbital energy (given by $E_{\rm orb}=-M_1 M_2/2b$)
to the rate of energy loss to infinity and to the holes:
\begin{equation}
{dE_{\rm orb}\over dt} = {M_1 M_2\over 2b^2}{db\over dt} = -\left({dE\over dt}
	\right)_N -{dM_1\over dt} - {dM_2\over dt},
\label{eq:dbdt}
\end{equation}
where $(dE/dt)_N$ is given in Eqs.~(\ref{eq:quadrupole}) and $\dot{M}_1$
in Eqs.~(\ref{eq:alignedspins}) [with $\dot{M}_2$ obtained by
exchanging the subscripts $1\leftrightarrow 2$ in Eqs.~(\ref{eq:alignedspins})].

The number of
gravitational-wave cycles $N_1$ emitted to infinity from initial time $t_i$
to final time $t_f$ (corresponding to separations $b_i$ and $b_f$) is given by
\begin{equation}
N = \int_{t_i}^{t_f} dt {\Omega_N\over\pi} = {1\over\pi}\int_{b_i}^{b_f}
	db {dt\over db}\left({M\over b^3}\right)^{1/2},
\end{equation}
where $dt/db$ is determined from Eq.~(\ref{eq:dbdt}).  This number is to be
compared with the number of cycles $N_2$ obtained by ignoring black hole
absorption/emission of energy and angular momentum, i.e.\ by setting
$\dot{E}_{\rm orb}$ equal to $-(dE/dt)_N$.  The difference
$\Delta N=N_1-N_2$ measures the effect of
black hole absorption/emission on the binary's orbital evolution.

The values of $\Delta N$ obtained by setting $b_i$ to be the separation at which
the gravitational-wave frequency is 10\,Hz (the low-frequency end of the LIGO
band), $\chi_1$ and $\chi_2$
to be 0.998, and the spins to be aligned with ${\bf L}_N$ (i.e.
$\hat{\bf L}_N\cdot\hat{\bf S}_1=\hat{\bf L}_N\cdot\hat{\bf S}_2=1$) are displayed
in Table~\ref{table:cycles} for various choices of total mass $M$ (in units of
a solar mass $M_{\odot}$) and mass
ratio $M_1/M_2$.  In the table, the numbers without parentheses are obtained
by setting $b_f=6M$, and those with parentheses by setting $b_f$ to be the larger
of $2M$ or the separation at which the wave frequency is 1000\,Hz (the
high-frequency end of the LIGO band).
For non-rotating black holes ($\chi_1=\chi_2=0$), the
corresponding values of $\Delta N$ (with $b_f=6M$) are all less than $10^{-2}$.

\begin{table}
\begin{center}
\begin{tabular}{cccc}
$M (M_{\odot})$ & $M_1/M_2=1$ & $M_1/M_2=2$ & $M_1/M_2=4$ \\ \hline
5 & 0.07 (0.07) & 0.11 (0.11) & 0.23 (0.24) \\
20 & 0.05 (0.07) & 0.07 (0.10) & 0.16 (0.22) \\
50 & 0.03 (0.06) & 0.05 (0.08) & 0.11 (0.18) \\
\end{tabular}
\end{center}
\caption{Change $\Delta N$ in the number of gravitational-wave cycles due to
black hole absorption/emission, for various values of total mass $M$ and mass ratio
$M_1/M_2$.  The initial separation is such that the wave frequency is 10\,Hz
and the spins satisfy $\chi_1=\chi_2=0.998$ and
$\hat{\bf L}_N\cdot\hat{\bf S}_1=\hat{\bf L}_N\cdot\hat{\bf S}_2=1$.  The numbers
without parentheses are for a final separation $b_f$ of $6M$; those with
parentheses are for $b_f$ equal to the larger of $2M$ or the separation at which
the wave frequency is 1000\,Hz.}
\label{table:cycles}
\end{table}

The values of $\Delta N$ in Table~\ref{table:cycles} indicate that
black hole absorption/emission
of energy and angular momentum during inspiral
may not be an important effect for the detection (by LIGO and VIRGO) and
analysis of gravitational waves from comparable-mass black holes.
Indeed, post-Newtonian corrections to the equations of motion and energy
loss have far greater influence on the number of wave cycles emitted by the
binary \cite{bdiww,kidder}.  It should be noted, however, that black hole
absorption/emission could have a much larger impact on the orbital evolution
of rapidly rotating holes when they are
beginning to merge, as suggested by Price and Whelan \cite{p&w}.  They have presented
models in which the tidal torque that results from black hole absorption/emission
of angular momentum plays a crucial role in the late stages of binary evolution (see
Fig.~1 in \cite{p&w}).  The perturbative methods used in this paper (based on wide
separation of the binary) are not valid in the close limit analyzed in \cite{p&w}.

It has been pointed out by Hughes \cite{hughes} that in the extreme-mass-ratio limit,
black hole absorption/emission can strongly influence the binary's
orbital evolution and is an important effect for LISA.

\section{Non-equatorial orbits}
\label{sec:nonequatorial}

In general, binary black holes are not expected to have spins aligned with the orbital
angular momentum ${\bf L}_N$.  This misalignment causes the spins and orbit
to precess in a
complicated way due to spin-orbit and spin-spin coupling \cite{apost,kidder}.
Each black hole's companion is not
in general confined to the hole's equatorial plane, and so the formulas in
the previous section are not applicable.  However, for orbits suitably close
to the equatorial plane (see below for details),
one can imagine using an approximation scheme in which
at each instant the companion's $\theta$-velocity is ignored; that is, the
companion is taken to be rigidly rotating in the $\phi$-direction at each point
on the orbit.  The changes in black hole parameters can then be calculated by
putting the instantaneous $\phi$-velocity into the rigid $\phi$-rotation
formulas~(\ref{eq:rigidrot}) at each point on the orbit.  In this section,
I construct such an approximation scheme.

\subsection{Description of orbit}

The evolutions of the spins and orbit are described by the equations
\cite{apost,kidder}
\begin{equation}
\dot{{\bf S}}_B = {\bf\Omega}^{(B)}_{\rm spin}\times{\bf S}_B,
	\qquad \dot{{\bf L}}_N = {\bf\Omega}_{\rm orb}\times{\bf L}_N
	-{32\over 5}\eta^2 Mv^7\hat{{\bf L}}_N,
\end{equation}
for $B=1,2$.
The orders of magnitude of the precession frequencies are
$\Omega_{\rm spin}\sim v^3 b^{-1}$ and $\Omega_{\rm orb}\sim v^4 b^{-1}$.  Since
the Newtonian angular velocity is $\Omega_N\sim vb^{-1}$, both $\Omega_{\rm spin}$
and $\Omega_{\rm orb}$ are much smaller than $\Omega_N$.
This means that, over a few orbital periods, ${\bf L}_N(t)$ and ${\bf S}_B(t)$ do not
change much due to precession.  Thus, the companion's orbit as seen in LARF1
is to a good approximation confined to a single plane with normal vector
${\bf n}=\hat{{\bf L}}_N(t)$ along the instantaneous direction of the
orbital angular momentum, on timescales of a few orbital periods.

In this subsection, I analyze
the trajectory of a particle in a planar, circular orbit of arbitrary
orientation in a fictitious Euclidean 3-space in terms of spherical
coordinates.  This information will be used to specify the rotation
rate and orientation of the companion's tidal field as seen in LARF1.
Denote the particle's radial coordinate by $b$, its constant (non-negative)
angular velocity by $\omega$, and the normal to its
orbital plane by ${\bf n}$.  The angle of inclination of the normal
with respect to the {\it z}-axis is denoted $\theta_n$; so
$\cos\theta_n={\bf n}\cdot{\bf e}_{\hat{z}}=n_z$.  Assume the orbit is centered on
the origin, so the particle's position ${\bf X}(t)$ at time $t$ is given
by a rotation $R({\bf n},\omega t)$ about ${\bf n}$, by an angle $\omega t$,
of the initial position ${\bf X}_0$.

In Cartesian coordinates, the particle's trajectory is given by
${\bf X}(t)={\bf X}_0\cos\omega t+\mbox{$({\bf n}\times{\bf X}_0)$}\sin\omega t$.
In terms of the particle's angular
coordinates $\theta(t)$ and $\phi(t)$, ${\bf X}(t)$ is equal to
$b[\sin\theta(t)\cos\phi(t),\sin\theta(t)\sin\phi(t),\cos\theta(t)]$.
I choose the initial position to be in the equatorial plane, i.e.
$Z_0={\bf X}_0\cdot{\bf e}_{\hat{z}}=0$.  This choice does not affect the
orbit-averaged quantities I calculate later in this section.

The angular functions $\theta(t)$ and $\phi(t)$ can now be expressed
in terms of ${\bf n}$ and $\omega$ using the relations above.  The
quantities of interest are $\sin^2\theta(t)$ and $\dot{\phi}(t)$, which are
determined to be
\begin{equation}
\sin^2\theta(t) = 1-\sin^2\theta_n\sin^2\omega t, \qquad
	\dot{\phi}(t) = \frac{\omega\cos\theta_n}{\sin^2\theta(t)}.
\label{eq:phidot}
\end{equation}

\subsection{Approximation scheme}

Return now to our black hole binary, and go to Boyer-Lindquist coordinates
$(t,r,\theta,\phi)$ in LARF1.  The companion's trajectory as seen in LARF1
will be described (to lowest order in $v$) by angular functions
$\theta(t)$ and $\phi(t)$ given by the expressions~(\ref{eq:phidot})
with $\omega$ replaced
by $\Omega_N$ and $\theta_n$ now referring to the angle of inclination of 
${\bf L}_N(t)$ with respect to ${\bf S}_1(t)$, that is, $\cos\theta_n
=\hat{\bf L}_N(t)\cdot\hat{\bf S}_1(t)$.  After these substitutions, we have
\begin{equation}
\sin^2\theta(t) = 1-\left(1-\left[\hat{\bf L}_N(t)\cdot\hat{\bf S}_1(t)\right]^2
	\right)\sin^2\Omega_N t,\qquad
	\dot{\phi}(t) = \frac{\Omega_N\hat{\bf L}_N(t)\cdot\hat{\bf S}_1(t)}
	{\sin^2\theta(t)}.
\label{eq:larfphidot}
\end{equation}
Since $\theta_n$ is now time dependent, these expressions are meaningful only
when used to calculate orbit-averaged quantities.

Consider the regime in which $\sin^2\theta(t)$ and $\dot{\phi}(t)$
are slowly varying; more precisely, require them to be approximately
constant on the timescale $\kappa_1^{-1}$ associated with the horizon.
As noted before, this is the timescale on which the null generators
of the horizon probe the future \cite{h&h,carter,membrane}.  The teleological
behavior of the horizon is, however, exponentially limited;  that is,
the influence of future events on the horizon decays exponentially in time,
with decay rate $\kappa_1$ (see, e.g., the discussion of
teleological Green functions in \cite{membrane}).  We thus
require $\Bigl|\dot{\phi}\Big/\ddot{\phi}\Bigr|$ and
$\left|(1/\sin^2\theta)d(\sin^2\theta)/dt\right|^{-1}$
[which are the same to leading order by Eqs.~(\ref{eq:larfphidot})]
to be only several times larger than $\kappa_1^{-1}$, rather than orders of magnitude
larger.

By assumption, $\chi_1\leq 0.998$, so $\kappa_1^{-1}$ is less than $34 M_1$.
Our requirement can then be expressed as
\begin{equation}
34\alpha M_1\Omega_N\leq\frac{\sin^2\theta(t)}{\sin^2\theta_n|\sin 2\Omega_N t|}
\label{eq:constraint}
\end{equation}
for all $t$, where $\alpha$ is a number roughly in the range 2-4.
A sufficient condition for this constraint to be satisfied
is $\cot^2\theta_n\geq 34\alpha M_1\Omega_N$.
This requires $\hat{{\bf L}}_N(t)$ to be near
one of the polar axes $\pm\hat{{\bf S}}_1(t)$, which correspond to $\theta=0,\pi$;
or, equivalently, the orbital plane must be near the equatorial plane.

We are interested in separations as small as $b=6M$, so $\Omega_N$ can be as large as
$(6^{3/2}M)^{-1}$.  For this reason, I impose the constraint $\cot^2\theta_n\geq 34\alpha
6^{-3/2}$ and set $\alpha$ to be approximately 3, obtaining the approximate constraints
$0\leq\theta_n\lesssim\pi/9$ or $8\pi/9\lesssim\theta_n\leq\pi$.  In other words,
$\hat{{\bf L}}_N(t)$ is within 20-degree cones around the polar axes, or, equivalently,
the inclination angle of the orbit with respect to the equatorial plane
is less than or (approx.) equal to 20 degrees.

For the approximation scheme in this section to be valid, we require further that
in the horizon's reference frame, the external tidal field should rotate primarily
in the $\phi$-direction and not significantly in the $\theta$-direction.  More
precisely, we require $\left|\dot{\theta}\right|\ll\left|\dot{\phi}-\Omega_{H1}\right|$.
The rates of change presented in Eqs.~(\ref{eq:nonequatorial}) and
(\ref{eq:orbitaverages}) below are subject to this condition.  For most values of
$\chi_1$, this condition is automatically satisfied throughout inspiral
(down to $b=6M$).  Even
if it is not satisfied at some point during inspiral, the restriction on $\theta_n$
discussed above ensures that the effect of the $\theta$-motion, when integrated over
inspiral, is negligible compared to that of the $\phi$-motion, for almost all values
of $\chi_1$.

With the above restriction on $\theta_n$, we can at each instant take $\sin^2\theta(t)$
and $\dot{\phi}(t)$ to be constant relative to the horizon timescale $\kappa_1^{-1}$,
and apply the rigid $\phi$-rotation
formulas~(\ref{eq:rigidrot}) with the instantaneous values $\theta(t)$
and $\dot{\phi}(t)$ put in.  This yields
\begin{eqnarray}
{dS_1\over dt} &=& \left[\dot{\phi}(t)-\Omega_{H1}\right]I_0[\theta(t)],
	\qquad {dM_1\over dt} = \dot{\phi}(t)\left[\dot{\phi}(t)-\Omega_{H1}
	\right]I_0[\theta(t)],\nn\\
\frac{\kappa_1}{8\pi}{dA_1\over dt} &=&
	\left[\dot{\phi}(t)-\Omega_{H1}\right]^2 I_0[\theta(t)],\nn\\
\label{eq:nonequatorial}
\end{eqnarray}
where $\theta(t)$ and $\dot{\phi}(t)$ are given by
Eqs.~(\ref{eq:larfphidot}) and $I_0$ by Eq.~(\ref{eq:I0}).

\subsection{Orbit-averaged quantities}

Next I would like to average these rates of change over an orbit assuming
the binary's masses, spins, separation, and orbital angular momentum
are approximately constant over an orbital period.  This assumption is
justified by the following ordering of the relevant timescales:
$\Omega_N^{-1}\ll\Omega_{\rm spin}^{-1},\Omega_{\rm orb}^{-1}\ll
\tau_{\rm ins}\ll\tau_S\ll\tau_M$.  We can therefore take
all the quantities on the right-hand sides of
Eqs.~(\ref{eq:nonequatorial}) except $\theta(t)$ and $\dot{\phi}(t)$
to be constant, to a good approximation, when averaging over an orbit.
Denote orbit averages
by angular brackets $\langle\rangle$; so, for example, $\langle \dot{S}_1
\rangle = (\Omega_N/2\pi)\int_0^{2\pi/\Omega_N}\dot{S}_1 dt$.
Plugging the expressions~(\ref{eq:larfphidot}) into Eqs.~(\ref{eq:nonequatorial})
and performing the orbit averages (as defined above) yields
\begin{eqnarray}
\left\langle\frac{dS_1}{dt}\right\rangle &=& \frac{r_{H1}}{10b^6}M_1^5 M_2^2\biggl(
	16(1+3\chi_1^2)\Bigl\{2\Omega_N{\cal N}_1(t)-\Omega_{H1}\left[{\cal N}_1^2(t)+1
	\right]\Bigr\}\nn\\
	&&\mbox{} +15\chi_1^2\left[
	{\cal N}_1^2(t)-1\right]\Bigl\{4\Omega_N{\cal N}_1(t)-\Omega_{H1}
	\left[3{\cal N}_1^2(t)+1\right]\Bigr\}\biggr),\nn\\
\left\langle\frac{dM_1}{dt}\right\rangle &=& \frac{2r_{H1}}{5b^6}M_1^5 M_2^2\Omega_N
	{\cal N}_1(t)\biggl(2\Omega_N(4-3\chi_1^2){\rm sign}[{\cal N}_1(t)]
	-8\Omega_{H1}(1+3\chi_1^2)\nn\\
	&&\mbox{} +15\chi_1^2\Bigl\{2\Omega_N{\cal N}_1(t)+\Omega_{H1}
	\left[1-{\cal N}_1^2(t)\right]\Bigr\}\biggr),\nn\\
\left\langle\frac{dA_1}{dt}\right\rangle &=& \frac{8\pi r_{H1}^2 M_1^5 M_2^2}
	{5b^6\left(1-\chi_1^2\right)^{1/2}}\biggl(
	(16-12\chi_1^2)\Bigl\{\Omega_{H1}^2\left[{\cal N}_1^2(t)+1\right]
	-4\Omega_{H1}\Omega_N{\cal N}_1(t)\nn\\
	&&\mbox{} +2\Omega_N^2\left|{\cal N}_1(t)\right|\Bigr\}
	+15\chi_1^2\Bigl\{\Omega_{H1}^2\left[3{\cal N}_1^4(t)+2{\cal N}_1^2(t)
	+3\right]\nn\\
	&&\mbox{} -8\Omega_{H1}\Omega_N{\cal N}_1(t)\left[{\cal N}_1^2(t)+1\right]
	+8\Omega_N^2{\cal N}_1^2(t)\Bigr\}\biggr),\nn\\
\label{eq:orbitaverages}
\end{eqnarray}
where ${\cal N}_B(t)=\hat{\bf L}_N(t)\cdot\hat{\bf S}_B(t)$ for $B=1,2$.
The corresponding expressions for BH2 can be obtained by exchanging the subscripts
$1\leftrightarrow 2$ in Eqs.~(\ref{eq:orbitaverages}).  Note that these
equations are valid only for ${\cal N}_B(t)$ suitably close to $\pm 1$,
as discussed above.  The formulas~(\ref{eq:orbitaverages}) can be applied
to a black hole in a binary with any companion body (e.g.\ a neutron
star) that has mass $M_2$ and is well separated from the hole.  

Numerical integration of Eqs.~(\ref{eq:orbitaverages}) using the
2.5 post-Newtonian equations of motion for spinning bodies
(\cite{kidder} and references therein) yields results comparable to those in
Tables~\ref{table:spin}-\ref{table:area}.

\section{Discussion}
\label{sec:discussion}

Having obtained the leading-order changes in a black hole's mass and spin during
inspiral [see Eqs.~(\ref{eq:changes})], we must check whether these changes
exceed the ambiguities inherent in the definitions of mass and spin
\cite{t&h}.  Denote by $\delta M$ and
$\delta S$ the magnitudes of the mass and spin ambiguities.  From
Eqs.~(1.8) in \cite{t&h},
\begin{equation}
\delta M \sim {ML^2\over {\cal R}^2}, \qquad \delta S\sim {M^3 L\over{\cal R}^2},
\end{equation}
where $M$ and $L$ are the mass and size of the (isolated) body in question,
and ${\cal R}$ is the external universe's radius of curvature.  For a black hole in
a binary, say BH1, $L\sim M_1$ and ${\cal R}^2\sim b^3/M_2$.  This implies
\begin{equation}
{\delta M_1\over M_1}\sim{\delta S_1\over M_1^2}\sim\eta{M_1\over M}\left(
	{M\over b}\right)^3.
\end{equation}

From Eqs.~(\ref{eq:changes}), the changes $\Delta M_1$ and $\Delta S_1$
from infinite separation to separation $b$ are
\begin{equation}
{\Delta M_1\over M_1}\sim\eta\left({M_1\over M}\right)^2\left(
	{M\over b}\right)^{7/2},\qquad {\Delta S_1\over M_1^2}\sim\eta{M_1\over M}
	\left({M\over b}\right)^2.
\end{equation}
So, at separation $b$, we have
\begin{equation}
{\Delta M_1\over\delta M_1}\sim{M_1\over M}\left({M\over b}\right)^{1/2},\qquad
	{\Delta S_1\over\delta S_1}\sim{b\over M}.
\end{equation}
We conclude that $|\Delta S_1|$ exceeds the ambiguity $\delta S_1$ in the
definition of spin, but $|\Delta M_1|$ does not rise above $\delta M_1$.
Note that the concept of tidal work is unambiguous \cite{purdue,b&c,favata}.

When analyzing initial data that contain a black hole and
represent the interface between inspiral and merger,
one can define and calculate the hole's mass and spin in
different ways, giving different answers corresponding to the ambiguities
$\delta M$ and $\delta S$ discussed above.  Since $\delta M$ is larger than
$|\Delta M|$, the hole's mass can be considered constant during
inspiral to the same level of accuracy as used in defining mass.  On the other hand,
$|\Delta S|$ exceeds $\delta S$, so the hole's spin cannot be
considered constant; however, as Table~\ref{table:spin} indicates,
the changes in spin are small during inspiral.

The results of this work---in particular, Eqs.~(\ref{eq:changes}) and
(\ref{eq:orbitaverages})---can be used
to relate the spin and horizon area of a black hole in a particular initial
data set to the spin and horizon area the hole had when infinitely separated from its
companion.

\section*{Acknowledgments}

I am grateful to Kip Thorne for helpful discussions and advice, and to Scott Hughes
for useful comments on the manuscript.
This research was supported in part by NSF grant PHY-0099568 and NASA grant NAG5-10707.

\end{document}